\documentclass[conference]{IEEEtran}
\IEEEoverridecommandlockouts
\usepackage[noadjust]{cite}
\usepackage{amsmath, graphicx, amssymb, amsfonts, color, subcaption, amsthm, multirow, array, adjustbox, booktabs, lipsum, float}

\usepackage{soul}
\usepackage{algorithmic}
\usepackage{textcomp}
\usepackage{xcolor}
\def\BibTeX{{\rm B\kern-.05em{\sc i\kern-.025em b}\kern-.08em
    T\kern-.1667em\lower.7ex\hbox{E}\kern-.125emX}}
\begin{document}

\title{LoRa Modulation for Split Learning
\thanks{This work is part of the project IRENE (PID2020-115323RB-C31), funded by MCIN/AEI/10.13039/501100011033 and supported by the Catalan government through the project SGR-Cat 2021-01207.}
}

\author{\IEEEauthorblockN{Marc Martinez-Gost\IEEEauthorrefmark{1}\IEEEauthorrefmark{2}, Ana Pérez-Neira\IEEEauthorrefmark{1}\IEEEauthorrefmark{2}\IEEEauthorrefmark{3}, Miguel Ángel Lagunas\IEEEauthorrefmark{2}}
\IEEEauthorblockA{
\IEEEauthorrefmark{1}Centre Tecnològic de Telecomunicacions de Catalunya, Spain\\
\IEEEauthorrefmark{2}Dept. of Signal Theory and Communications, Universitat Politècnica de Catalunya, Spain\\
\IEEEauthorrefmark{3}ICREA Acadèmia, Spain\\
\{mmartinez, aperez, malagunas\}@cttc.es
}}

\maketitle

\begin{abstract}

In this paper we introduce a task-oriented communication design for split learning (SL) over a communication channel. Our approach involves the Expressive Neural Network (ENN), a novel neural network featuring adaptive activation functions (AAF) based on the Discrete Cosine Transform (DCT). This architecture does not only provide better learning capabilities, but also facilitates data transmission using the Long Range (LoRa) modulation. The frequency nature of LoRa is adequate for the communication side of the problem, while allowing to construct the AAFs at the receiver. Additionally, we propose orthogonal chirp division multiplexing (OCDM) for multiple access and a modified modulation aimed at preserving communication bandwidth. Our experimental results demonstrate the effectiveness of this scheme, achieving high accuracy in challenging scenarios, including low signal to noise Ratio (SNR) and absence of channel state information (CSI) for both additive white Gaussian noise (AWGN) and Rayleigh fading channels.
%This innovative approach holds promise for enhancing the efficiency and robustness of task-oriented communications in SL applications.
%Our results show that scheme achieves high accuracy up to -15 dB in the presence of additive white Gaussian noise (AWGN), and up to -12.5 dB in the case of Rayleigh fading.
\end{abstract}

\section{Introduction}
\label{sec:intro}
Split learning (SL) is a distributed machine learning technique that is suited for devices with limited computation capabilities \cite{gup18,vep18_2}. The sequential learning model, usually a neural network, is split between two (or more) devices.
%Since the data is processed at each device and only the intermediate computations are shared among devices, SL is a privacy-preserving technique.
Although there are many approaches to split the model \cite{vep18}, the most relevant trade-off in SL is at which location the split takes place: allocating less computations to one device, leaves most of the computational burden to the other one.

Splitting the network between several devices requires the implementation of a communication protocol between them, both during training and inference. In \cite{sin19} the authors study the communication efficiency in terms of the number of trainable parameters and the number of clients. For resource-constrained devices, in \cite{cho21} an alternative architecture is proposed to reduce the communication and computing cost during training, while in \cite{che21} the number of updates is limited and the data quantized. Focusing on the resource allocation, the devices in \cite{Wu23} are clustered to speed up convergence and reduce the communication cost.
Despite the notorious interest in distributed learning frameworks, particularly in federated learning, SL has not received that much attention in the context of task-oriented communications \cite{sahin22, Gunduz23}.
In \cite{bou19} a deep  joint source-channel code is proposed for wireless image transmission. The neural network is split and trained to map the image pixel values to the complex-valued channel input symbols.
However, there is no literature devoted to the design of the physical layer for SL, this is, specific waveforms and multiple access schemes to deploy a neural network over wireless communications.

In this work we first propose a neural network termed Expressive Neural Network (ENN) that is suitable for SL. The model features adaptive activation functions (AAF) shaped with the Discrete Cosine Transform (DCT) and whose coefficients are learnt during backpropagation.
Secondly, we suggest a novel design for task-oriented communications. With an appropriate split of ENN, the transmitter can send the information using the Long Range (LoRa) modulation. This is because the DCT characterization to generate the AAFs at the receiver match the frequency nature of LoRa.
Both the splitting model and the modulation are suited for energy-constrained devices.
While in our previous works \cite{mar23_2, mar23_3} we show the benefits of this modulation for communication, in the present work we demonstrate that it also assists the computing side of the problem
%via DCT-modeled AAFs.
Thus, the DCT is used to characterize the non-linearities, but also to propagate the information throughout the communication channel. An extensive description and analysis of ENN can be found in \cite{mar23}, where we show the generalization capabilities of the model and the expressiveness that the DCT provides.

%Following a signal processing perspective, we propose a novel design for task-oriented communications, in which the task is training a neural network in a SL fashion. Both the splitting model and the modulation are suited for energy-constrained devices. 
Furthermore, we also provide an orthogonal chirp division multiplexing (OCDM) for multiple access. This relies on a chirp spread spectrum (CSS) technique, which is also implemented in LoRa. We provide extensive results for different communication channels, demonstrating that LoRa in SL provides high accuracy even at low signal to noise ratio (SNR) and requires no channel state information (CSI). Finally, we also propose a modified version of the modulation that allows to reduce the transmission bandwidth. %with respect to the original modulation.

\section{Expressive Neural Network (ENN)}

%Consider a scalar univariate function $f(x)$ to be approximated over the input variable range $x\in[-1,1]$. The Discrete Cosine Transform (DCT) has the following expression:
%\begin{equation}
%f(x) \approx \sum_{q=1}^{Q} g_qF_q \cos\left(\frac{\pi (q-1)(2z+1)}{2N}\right),
%\label{eq:idct}
%\end{equation}
%with $z=\frac{N}{2}(1+x)$, $g_1=1/\sqrt N$ and $g_q=\sqrt{2/N}$ otherwise. The $F_q\in\mathbb{R}$ are termed the DCT coefficients. Notice that for functions with odd symmetry, only the odd coefficients are retained. In general, the quality of approximation is more than sufficient for $Q=12$ (i.e., 6 coefficients in odd functions). For the sake of brevity, the following definition will be used when needed:

%In \cite{Per23} we propose an adaptive design in which the DCT coefficients of \eqref{eq:idct} are tuned using the Least Mean Squares (LMS) algorithm in a supervised setting to approximate an univariate function.

%\mmg{This architecture, termed ENN, will be fully-adaptive as both the linear weights and the activation functions will be trained in a supervised fashion.}

\begin{figure}[t]
    \centering
    \includegraphics[width=\columnwidth]{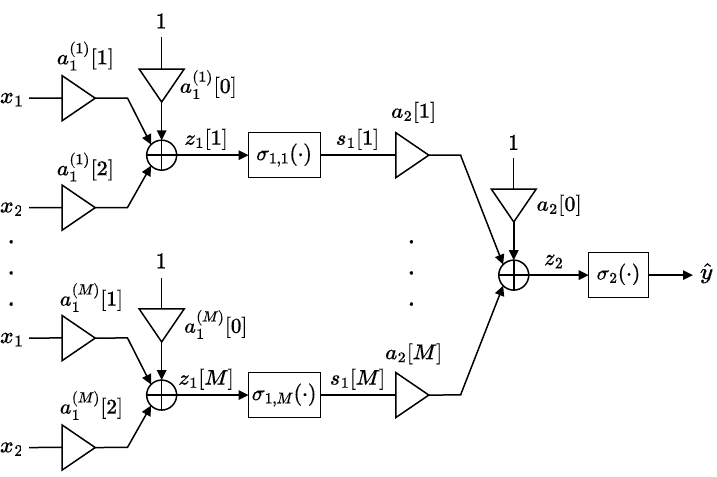}
  \caption{A 2-layer perceptron with $M$ neurons in the hidden layer.}
  \label{fig:2_layer_perceptron}
  \vspace{-0pt}
\end{figure}

Consider the 2-layer neural network shown in Fig. \ref{fig:2_layer_perceptron}, consisting of a two-input vector $\mathbf{x}=[x_1,x_2]$ and a hidden layer of $M$ neurons (or perceptrons).
A neuron is a non-linear processor involving a weighted sum and a non-linear function. The following expressions show the perceptron signals at the $l$-th layer ($l=1,2$):
\begin{align}
    \mathbf{z}_l &= \mathbf{A}_l^T[1\,\, \mathbf{s}_{l-1}^T]^T
    \label{eq:linear_multi}\\
    %\overline{\mathbf{z}}_l &= \frac{N}{2}(\mathbf{z}_l+1)
    %\label{eq:normalization_multi}\\
    \mathbf{s}_l &= \sigma_{l,m} ({\mathbf{z}}_l)
    \label{eq:non_linearity_multi}
\end{align}
The first input corresponds to $\mathbf{s}_{0}=\mathbf{x}$, while the last output is the predicted class or regression value, namely $\mathbf{s}_L=\hat{y}$. The matrix of linear weights is $\mathbf{A}_l=[\mathbf{a}_l^{(1)}\,\dots\,\mathbf{a}_l^{(M)}]$. The first entry of $\mathbf{s}_{l-1}$ is $\mathbf{s}_{l-1}[0]=1$, associated to the bias term $\mathbf{a}_{l}^{(m)}[0]$.

%The normalization step in \eqref{eq:normalization_multi} is needed to map the input to $[0,N]$, assuming the input is confined to $[-1,1]$. Nevertheless, as it will be seen later, the first linear transformation may map $\overline{z}$ to a different range, providing expressiveness to the network.
The operation in \eqref{eq:non_linearity_multi} is performed element-wise, and the AAF $\sigma(\cdot)$ is approximated by the DCT with $Q/2$ coefficients, this is, the $m$-th element of \eqref{eq:non_linearity_multi} is computed as
\begin{equation}
    \mathbf{s}_l[m] =
    %\sum_{q=1}^{Q/2} F_{q,l}^{(m)}\cos\left(\frac{\pi (2q-1)(2\overline{\mathbf{z}}_l[m]+1)}{2N}\right)
    \sum_{q=1}^{Q/2} F_{q,l}^{(m)}\cos_q(\mathbf{z}_l[m]),
    \label{eq:non_linearity_neuron}
\end{equation}
with
\begin{equation}
    cos_i(x) = \cos\left(
\frac{\pi}{2N}\left(2i-1\right)\left(N(x+1)+1
\right)
\right),
\label{eq:cosine_definition}
\end{equation}
where $F_{q,l}^{(m)}$ corresponds to the $q$-th coefficient of the $m$-th perceptron at the $l$-th layer. As explicitly shown in \eqref{eq:non_linearity_multi} and \eqref{eq:non_linearity_neuron}, the activation function is not necessarily the same at each neuron, although the number of DCT coefficients $Q$ is kept constant for the whole network.
%Without loss of generality, $g_q$ is assumed to be integrated in the DCT coefficient $F_q$.
In \eqref{eq:non_linearity_neuron} we assume the AAF to have odd symmetry, so that only the odd coefficients are retained. Nevertheless, this does not prevent the network from learning only odd activation function.

There are numerous advantages in using this DCT representation: A small number of coefficients is required; a gradient-based adaptive algorithm can be implemented because the coefficients are real and ordered in decreasing magnitude;
since the basis functions (i.e., cosines) are orthogonal, the approximation error can be easily controlled by the magnitude of the disregarded coefficients, which also simplifies the convergence of the learning procedure;
furthermore, see that the index appears in the phase of \eqref{eq:non_linearity_neuron}, so the approximation is real and bounded, even when the input exceeds the dynamic range. All these features make the DCT an appropriate function approximation.
Besides, the number of parameters in ENN only increases by $Q/2$ with respect to a standard perceptron with fixed activation functions, while the expressiveness of the neural network increases dramatically and is general enough to be trained for both classification and regression problems.
See \cite{mar23} for a detailed description of ENN, the corresponding learning rules and an exhaustive list of experiments demonstrating its learning capabilities.

\section{Communication design for SL}
\label{sec:split_learnig}

\begin{figure*}[t]
    \centering
    \includegraphics[width=1\textwidth]{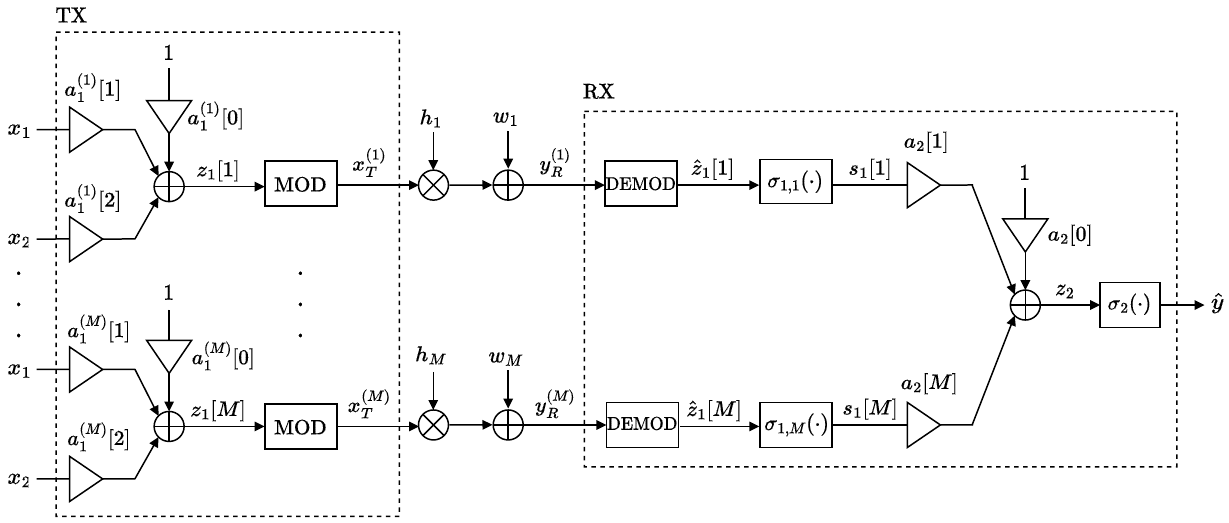}
  \caption{Splitting of the ENN between a transmitter (TX) and a receiver (RX) with a communication channel in-between.}
  \label{fig:ICASSP24_ENN_SL}
\end{figure*}

While there are many ways to split the network between a transmitter and a receiver, in the following we propose a scheme that is suited for both the communication and computing sides of the problem. Consider the split shown in Fig. \ref{fig:ICASSP24_ENN_SL}, where  the linear combinations of the first layer are left at the transmitter and the rest of the network is deployed at the receiver. Regarding the communication blocks, the output at the $m$-th neuron of the transmitter, $z_1[m]$, is modulated into $x_T^{(m)}$. This signal is sent through a channel and the discrete-time received signal is
\begin{equation}
    y_R^{(m)}[n] = h_mx_T^{(m)}[n] + w[n], \quad n=0,\dots,N-1,
    \label{eq: received_signal}
\end{equation}
where $h_m$ is the flat fading channel and $w[n]$ are the corresponding additive white Gaussian noise (AWGN) samples. The demodulator estimates $\hat{z}_1[m]$ over $y_R^{(m)}$ and proceeds with the following blocks of the ENN.

\subsection{Frequency modulation for SL}
Motivated by our previous work \cite{mar23_3}, we choose to modulate $z_1[m]$ in frequency as
\begin{equation}
    x_T^{(m)}[n]=A_c\sqrt{\frac{2}{N}}\cos\left(\frac{\pi(N(\overline{\mathbf{z}}_1[m]+1)+1)}{2N}n\right)
    \label{eq:modulation_v1}
\end{equation}
for $n=0,\dots,N-1$, where $A_c$ is the amplitude of the carrier and $\overline{\mathbf{z}}_1[m]$ is a quantized version of $\mathbf{z}_1[m]$ to the nearest integer. The $\sqrt{2/N}$ term is used to normalize the transmitted power. Notice that the waveform in \eqref{eq:modulation_v1} corresponds to the cosine in \eqref{eq:cosine_definition} for $i=1$ and with a discrete time index $n$. Therefore, this modulation is an $M$-ary Frequency Shift Keying ($M$-FSK), the one used in LoRa, but implementing the DCT and not the Discrete Fourier Transform (DFT) basis. LoRa is widely deployed in current wireless sensor networks \cite{Chi19}. From a computing perspective, this waveform already provides the cosine nature that the receiver will use to implement the non-linearities in \eqref{eq:non_linearity_neuron}. Nonetheless, the modulation is also advantageous from the communication side.
Since the waveforms in \eqref{eq:modulation_v1} correspond to the DCT, the demodulator computes the inverse DCT, and recovers a peak located at the corresponding frequency. The quantization at the transmitter allows to recover exactly one peak in the demodulation phase, with an associated error probability of
\begin{equation}
    P_e \approx (N-1)Q\left(\sqrt{\frac{A_c^2|h_m|^2}{N_o}}\right)
    \label{eq: error_prob_v1}
\end{equation}
As the amplitude does not carry information, the system is blind and no CSI is needed. On the other hand, the quantization noise is easily controlled by the number of samples $N$ and allows to implement a digital demodulation, which simplifies the demodulation phase with respect to a full analog scheme.
%Notice that \eqref{eq:modulation_v1} is no more than the Long Range (LoRa) modulation, which is widely deployed in current wireless sensor networks \cite{Chi19}.

Overall, the proposed split keeps a low computational load at the transmitter because only linear combinations are performed, and a low communication burden as this requires simple hardware (an analog-to-digital converter and a frequency modulator) and no CSI.

\subsection{Multiple access scheme}
\label{sec: ocdm}
As displayed in Fig. \ref{fig:ICASSP24_ENN_SL}, the transmission of $M$ data streams requires the use of $M$ different channel uses. Alternatively, there are techniques that prevent from using orthogonal resources, ranging from marking each stream with a different power to MIMO procedures. Motivated by our previous work \cite{mar23_2}, we propose to implement  OCDM \cite{ouy16}. It has been extensively shown the benefits of implementing CSS over LoRa signals \cite{Wu19}. In here we also use the $M$ orthogonal chirps to design the multiple access scheme.
Specifically, the following digital chirp,
\begin{equation}
    \psi_m[n] = e^{-j\frac{\pi}{N}\left(n-m\right)^2},\quad n=0,\dots,N-1,
\end{equation}
allows to generate $M$ orthogonal chirps for $m=0,\dots,M-1$ and for $N$ even. The resulting waveform is
\begin{equation}
    x_{T,CSS}^{(m)}[n] = x_T^{(m)}[n] e^{-j\frac{\pi}{N}\left(n-m\right)^2},\quad n=0,\dots,N-1
    \label{eq: ocdm_signal}
\end{equation}

These signals can be generated using a bank of filters, and in a similar fashion at the receiver side to recover each individual signal. The OCDM scheme increases the bandwidth proportional to the number of orthogonal chirps $M$.

% \begin{equation}
%     \mathbf{x}_{T,LoRa}^{(m)}= \mathbf{\Phi} \mathbf{x}_T^{(m)},
%     \label{eq: iDFnT}
% \end{equation}
% where $\mathbf{x}_{T}^{(m)}$ and $\mathbf{x}_{T,LoRa}^{(m)}$ are the vectorized version of $x_{T}^{(m)}[n]$ and $x_{T,LoRa}^{(m)}[n]$, and
% \begin{equation}
%     \mathbf{\Phi}(p,q) = \frac{1}{\sqrt{N}}e^{-j\frac{\pi}{4}}e^{-j\frac{\pi}{N}\left(p-q\right)^2}.
% \end{equation}

% OCDM corresponds to the inverse Discrete Fresnel Transform (iDFnT), which can be synthesised with bank of discretized modulated chirp as

% To recover the individual signals at the receiver, a bank of matched filters is implemented with the DFnT as
% \begin{equation}
%     \mathbf{x}_{T}^{(m)}= \mathbf{\Phi}^H \mathbf{x}_{T,LoRa}^{(m)}
%     \label{eq: DFnT}
% \end{equation}

%\vspace{-5pt}

\subsection{Gradient propagation}
While in Fig. \ref{fig:ICASSP24_ENN_SL} we show the architecture for propagating the data forward, the communication architecture also needs to be defined to propagate the gradient backwards and adapt the weights at the transmitter side. The gradient information used to update the $k$-th linear weight at the $m$-th neuron is
\begin{align}
    G(a_1^{(k)}[m])=&\frac{\pi^2}{4}\frac{s_0[m]}{|s_0[m]|^2}\varepsilon
     a_2[k]
    \sum_{p=1}^{Q/2}F_{p,2}(2p-1)\sin_p(z_2)\nonumber\\
    &\sum_{q=1}^{Q/2}F_{q,1}^ {(k)}(2q-1)\sin_q(z_1[k]),
    \label{eq:LMS_lin_1}
\end{align}
where $\varepsilon$ is the error of the learning task (see \cite{mar23} for a full description of the gradient expressions). Notice that this information is different for each $k$ and $m$, meaning that it also requires independent channel uses. Thus, we propose to use the same modulation and multiple access scheme from the forward pass in the backward pass: the information in $G$ (despite $s_0[m]$) is quantized and transmitted using the same LoRa modulation and OCDM.

\subsection{Constrained bandwidth}
The dynamic range of the modulated data is controlled in practice. At convergence, the linear weights of neural networks are distributed around zero and of small magnitude. However, there are no theoretical guarantees that this always happens and it is less certain during training. Furthermore, as shown in \cite{mar23}, the dynamic range of $z_1[m]$ exceeds $[0,N-1]$, which provides expressiveness to the architecture by exploiting the different periods of the DCT. Using the $M$-FSK modulation in \eqref{eq:modulation_v1}, this translates in having no bounds on the maximum transmitted frequency and, consequently, no control on the occupied bandwidth. 
While this generally may not represent an issue, in the following we propose an extension of the modulation that constraints $z_1[m]$ in the $[0,N-1]$ range and limits the maximum bandwidth. 

Due to the periodic extension of the DCT, it is easy to see the following relationship:
\begin{equation}
    \sigma(z) = (-1)^{\left\lfloor z/N\right\rfloor} \sigma(\bmod_N(|z|)),
\end{equation}
where $\lfloor\cdot\rfloor$ is the \textit{floor} operator and $\bmod_N(\cdot)$ is the modulo $N$ operator. For a given $z\in [kN,(k+1)(N-1)]$, the function value is the same as in the $[0,N-1]$ range when $k$ is even, whereas the sign is reversed for odd $k$. Thus, any point in the periodic extension of the function can be transposed to the original range. The corresponding modulation is
\begin{align}
    x&_T^{(m)}[n]=\nonumber\\ &A_c\sqrt{\frac{2}{N}}\cos\left(\frac{\pi(2\bmod_N(|\overline{\mathbf{z}}_1[m]|)+1)}{2N}n + \left\lfloor\frac{\overline{\mathbf{z}}_1[m]}{N}\right\rfloor\pi \right),
\end{align}
for $n=0,\dots,N-1$ and where the phase carries the range parity. This results in a joint frequency and phase modulation, where the latter corresponds to a Binary Phase Shift Keying (BSPK), i.e., the phase is either 0 or $\pi$. After estimating both parameters at the receiver, the non-linearity is implemented as
\begin{equation}
    \sigma_{1,m} = (-1)^{\left\lfloor \hat{\mathbf{z}}_1[m]/N\right\rfloor}
    \sum_{q=1}^{Q/2} F_{q,l}^{(m)}\cos_q(\hat{\mathbf{z}}_1[m]),
    \label{eq:non_linearity_v2}
\end{equation}

To find the bandwidth, we will use the continuous-time signal evaluated at the maximum frequency (i.e., $\overline{\mathbf{z}}_1[m]=N-1$). Considering a sampling frequency of $f_s=N/T$, where $T$ is the symbol period, this results in
\begin{equation}
    x_T^{(m)}(t) = A_c \cos\left(\frac{\pi(2N-1)}{2T}t\right)\approx
    A_c\cos\left(2\pi\frac{N}{2T}t\right)
\end{equation}

The bandwidth occupied by this modulation is $B=N/2T$. The price to pay for a reduced bandwidth is having a larger error probability in demodulation. Since the frequency and phase are independent, the total error corresponds to the demodulation error of $M$-FSK and BPSK:
\begin{equation}
    P_e \approx (N-1)Q\left(\sqrt{\frac{A_c^2|h_m|^2}{N_o}}\right) + Q\left(\sqrt{\frac{2A_c^2|h_m|^2}{N_o}}\right),
    \label{eq: error_prob}
\end{equation}
in which the first term dominates. While implementing this joint frequency and phase modulation does not really increase the error probability, the modulation now requires CSI, at least to compensate the phase introduced by the channel. Thus, an intrinsic trade-off appears: reducing the transmission bandwidth requires implementing CSI.
This may be implemented at the receiver, which does not increase the complexity at the transmitters.

\section{Experimental results}
\label{sec:results}
% Rayleigh:
%     - no noise, only quant: 2.08
%     - 0 dB: 
%     - -5 dB: 
%     - -10 dB: 
%     - -12.5 dB: 
%     - -15 dB: 
%     - -17.5 dB: 
%     - -20 dB: 

\begin{figure}[t]
\centering
\vspace{0.1in}
     \begin{subfigure}[b]{0.23\columnwidth}
         \includegraphics[width=\columnwidth]{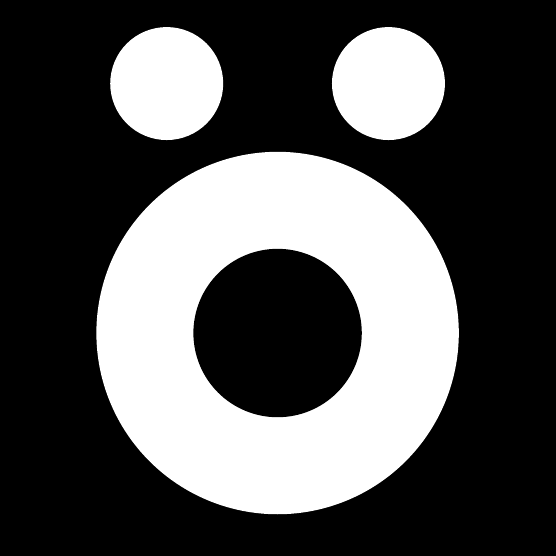}
         \caption[]{Ideal.} % <---
         \label{fig:map_face_ideal}
     \end{subfigure}
     \hspace{30pt}
     \begin{subfigure}[b]{0.23\columnwidth}
         \includegraphics[width=\columnwidth,angle=90]{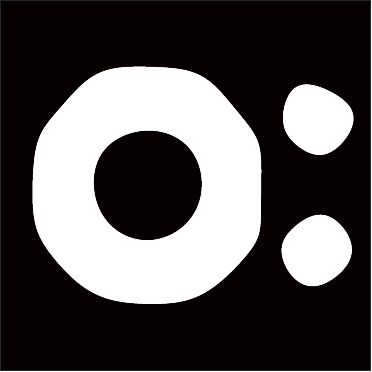}
         \caption[]{ENN.} % <---
         \label{fig:map_face_ENN}
     \end{subfigure}
     \hspace{0pt}%\hfill

     \caption{Decision map for the binary classification problem: (a) ideal case and (b) learnt with ENN (accuracy: $97.8\%$).}
     \label{fig:face_maps}
\end{figure}

The ENN is built with $M=6$ neurons in the hidden layer and with $Q/2=6$ parameters in all the AAFs. The ENN is trained with the Least Mean Squares (LMS) algorithm and the mean squared error (MSE) loss, following the same configuration as in \cite{mar23}. The ENN is trained for the binary classification problem shown in Fig. \ref{fig:map_face_ideal}. All samples come from independent uniform distributions in the $[-1,1]$ range for each input variable.
The train and test sets contain 800.000 and 50.000 samples, respectively.

We test the SL architecture in the ENN for different channel models, while the benchmark is the centralized ENN, in which there is no communication. Fig. \ref{fig:map_face_ENN} shows the decision boundary achieved by the benchmark, providing an accuracy of $97.8\%$. In all the scenarios we assume that the receiver has enough transmission power to work above an SNR of -10 dB, providing almost no errors in backpropagation. %when demodulating the gradients.

\begin{table}[t]
  \begin{center}
    \begin{tabular}{c|cccccc}
      \toprule % <-- Toprule here
      SNR (dB) & 0 & -5 & -10  & -12.5 & -15 & -17.5\\
      \midrule
     
      \multirow{3.7}{*}{AWGN} &
      $97.5\%$ &
      $97.4\%$ &
      $96.9\%$ &
      $95.0\%$ &
      $82.0\%$ &
      $71.9\%$
      %$67.6\%$
      \\
      
       &
       \adjustbox{valign=c}{\includegraphics[width=0.3in, angle=90]{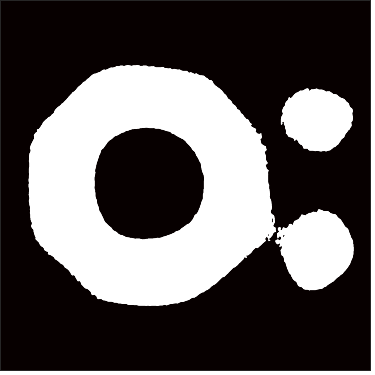}}
      &
      \adjustbox{valign=c}{\includegraphics[width=0.3in, angle=90]{img/ENN_AWGN_snr_-5.pdf}}
      &
      \adjustbox{valign=c}{\includegraphics[width=0.3in, angle=90]{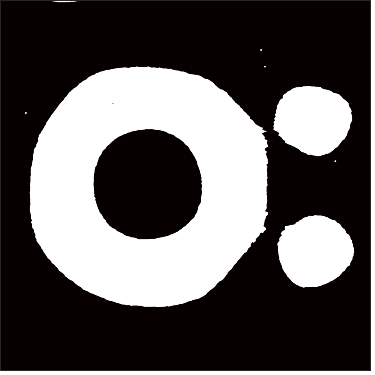}}
      &
      \adjustbox{valign=c}{\includegraphics[width=0.3in, angle=90]{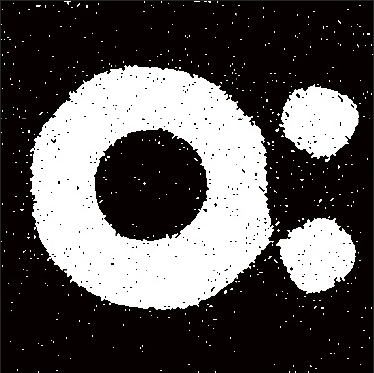}}
      &
      \adjustbox{valign=c}{\includegraphics[width=0.3in, angle=90]{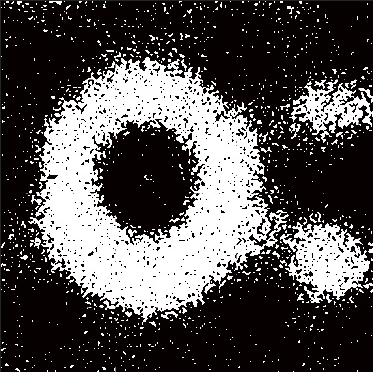}}
      & \adjustbox{valign=c}{\includegraphics[width=0.3in, angle=90]{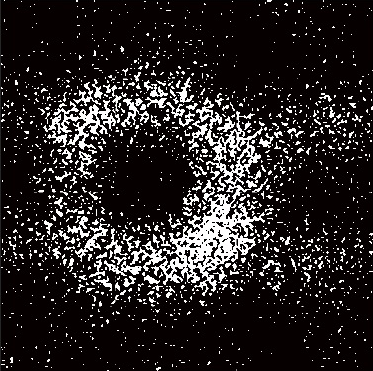}}
      % &
      % \adjustbox{valign=c}{\includegraphics[width=0.3in, angle=90]{img/ENN_AWGN_snr_-20.pdf}}
      \\
      \midrule

      \multirow{3.7}{*}{Rayleigh} &
      $96.2\%$ &
      $91.5\%$ &
      $86.8\%$ &
      $84.4\%$ &
      $79.5\%$ &
      $75.5\%$
      %$71.9\%$
      \\
      
      &
      \adjustbox{valign=c}{\includegraphics[width=0.3in, angle=90]{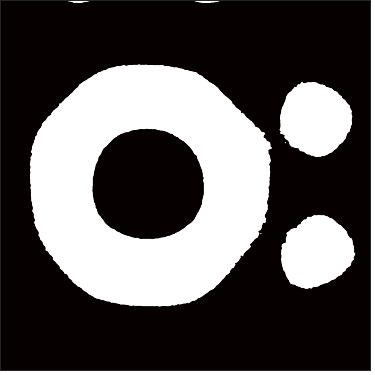}}
      &
      \adjustbox{valign=c}{\includegraphics[width=0.3in, angle=90]{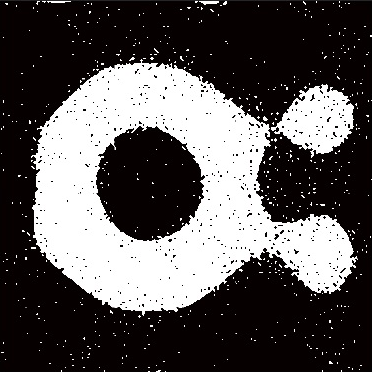}}
      & 
      \adjustbox{valign=c}{\includegraphics[width=0.3in, angle=90]{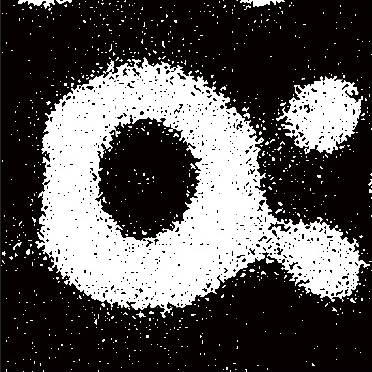}}
      & \adjustbox{valign=c}{\includegraphics[width=0.3in, angle=90]{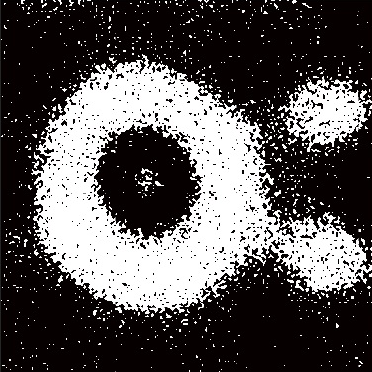}}
      & 
      \adjustbox{valign=c}{\includegraphics[width=0.3in, angle=90]{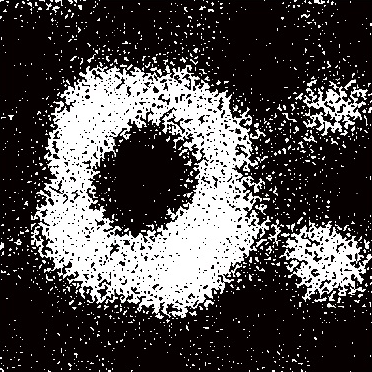}}
      & \adjustbox{valign=c}{\includegraphics[width=0.3in, angle=90]{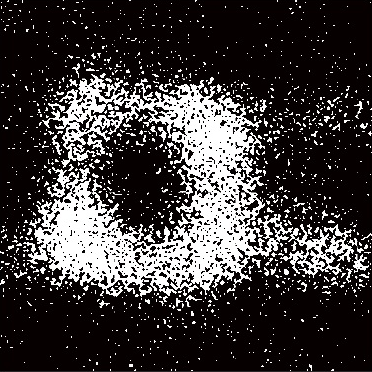}}
      % &
      % \adjustbox{valign=c}{\includegraphics[width=0.3in, angle=90]{img/ENN_Rayleigh_snr_-5.pdf}}
      \\
      
      \bottomrule
    \end{tabular}
    \caption{Accuracy and decision map for AWGN and Rayleigh channels at different SNR.}
    \label{tab:classification_results}
    \vspace{-10pt}
  \end{center}
\end{table}

Table \ref{tab:classification_results} shows accuracy and decision maps for the AWGN and Rayleigh channels at different SNR levels. In the AWGN case, above $-10$ dB there are no errors in demodulation, and the accuracy is almost as in the centralized ENN. Thus, the only source of errors is due to quantization, which is negligible. High accuracy is achieved above -15 dB in the AWGN channel and above -12.5 dB in the presence of Rayleigh fading.

%\mmg{Results on BW?}

\section{Conclusions}

In this paper we focus on task-oriented communications for SL over a communication channel. We have proposed a neural network model and a corresponding physical layer design. Specifically, ENN is an architecture with DCT-based adaptive activation functions. Besides improving its learning capabilities, the DCT representation allows to split the network and use the LoRa modulation to transmit the information between the splits. In this respect, the frequency nature of LoRa is suitable from a communication side, and also provides the characterization needed to construct the activation functions at the receiver. Furthermore, we propose an OCDM for multiple access and a variant of the modulation to preserve bandwidth.
Our results show that the  scheme provides high accuracy even for low SNR and without CSI for both AWGN and Rayleigh fading channels.

\bibliographystyle{IEEEbib}
\bibliography{refs}

\begin{thebibliography}{10}

\bibitem{gup18}
Otkrist Gupta and Ramesh Raskar,
\newblock ``Distributed learning of deep neural network over multiple agents,''
\newblock {\em Journal of Network and Computer Applications}, vol. 116, pp.
  1--8, 2018.

\bibitem{vep18_2}
Praneeth Vepakomma, Tristan Swedish, Ramesh Raskar, Otkrist Gupta, and
  Abhimanyu Dubey,
\newblock ``No peek: A survey of private distributed deep learning,''
\newblock {\em arXiv preprint arXiv:1812.03288}, 2018.

\bibitem{vep18}
Praneeth Vepakomma, Otkrist Gupta, Tristan Swedish, and Ramesh Raskar,
\newblock ``Split learning for health: Distributed deep learning without
  sharing raw patient data,''
\newblock {\em arXiv preprint arXiv:1812.00564}, 2018.

\bibitem{sin19}
Abhishek Singh, Praneeth Vepakomma, Otkrist Gupta, and Ramesh Raskar,
\newblock ``Detailed comparison of communication efficiency of split learning
  and federated learning,''
\newblock {\em arXiv preprint arXiv:1909.09145}, 2019.

\bibitem{cho21}
Ayush Chopra, Surya~Kant Sahu, Abhishek Singh, Abhinav Java, Praneeth
  Vepakomma, Vivek Sharma, and Ramesh Raskar,
\newblock ``Adasplit: Adaptive trade-offs for resource-constrained distributed
  deep learning,''
\newblock {\em arXiv preprint arXiv:2112.01637}, 2021.

\bibitem{che21}
Xing Chen, Jingtao Li, and Chaitali Chakrabarti,
\newblock ``Communication and computation reduction for split learning using
  asynchronous training,''
\newblock in {\em 2021 IEEE Workshop on Signal Processing Systems (SiPS)},
  2021, pp. 76--81.

\bibitem{Wu23}
Wen Wu, Mushu Li, Kaige Qu, Conghao Zhou, Xuemin Shen, Weihua Zhuang, Xu~Li,
  and Weisen Shi,
\newblock ``Split learning over wireless networks: Parallel design and resource
  management,''
\newblock {\em IEEE Journal on Selected Areas in Communications}, vol. 41, no.
  4, pp. 1051--1066, 2023.

\bibitem{sahin22}
Alphan Sahin and Rui Yang,
\newblock ``A survey on over-the-air computation,''
\newblock {\em arXiv preprint arXiv:2210.11350}, 2022.

\bibitem{Gunduz23}
Deniz Gündüz, Zhijin Qin, Inaki~Estella Aguerri, Harpreet~S. Dhillon, Zhaohui
  Yang, Aylin Yener, Kai~Kit Wong, and Chan-Byoung Chae,
\newblock ``Beyond transmitting bits: Context, semantics, and task-oriented
  communications,''
\newblock {\em IEEE Journal on Selected Areas in Communications}, vol. 41, no.
  1, pp. 5--41, 2023.

\bibitem{bou19}
Eirina Bourtsoulatze, David~Burth Kurka, and Deniz G{\"u}nd{\"u}z,
\newblock ``Deep joint source-channel coding for wireless image transmission,''
\newblock {\em IEEE Transactions on Cognitive Communications and Networking},
  vol. 5, no. 3, pp. 567--579, 2019.

\bibitem{mar23_2}
Marc~M. Gost, Ana Pérez-Neira, and Miguel~Ángel Lagunas,
\newblock ``{DCT}-based air interface design for function computation,''
\newblock {\em IEEE Open Journal of Signal Processing}, vol. 4, pp. 44--51,
  2023.

\bibitem{mar23_3}
Marc~M. Gost, Ana Pérez-Neira, and Miguel~Ángel Lagunas,
\newblock ``{LoRa}-based over-the-air computing for {Sat-IoT},''
\newblock in {\em 2023 IEEE 31st European Signal Processing Conference
  (EUSIPCO)}, 2023.

\bibitem{mar23}
Marc Martinez-Gost, Ana P{\'e}rez-Neira, and Miguel~{\'A}ngel Lagunas,
\newblock ``{ENN}: A neural network with {DCT}-adaptive activation functions,''
\newblock {\em arXiv preprint arXiv:2307.00673}, 2023.

\bibitem{Chi19}
Marco Chiani and Ahmed Elzanaty,
\newblock ``On the {LoRa} modulation for {IoT}: Waveform properties and
  spectral analysis,''
\newblock {\em IEEE Internet of Things Journal}, vol. 6, no. 5, pp. 8463--8470,
  2019.

\bibitem{ouy16}
Xing Ouyang and Jian Zhao,
\newblock ``Orthogonal chirp division multiplexing,''
\newblock {\em IEEE Transactions on Communications}, vol. 64, no. 9, pp.
  3946--3957, 2016.

\bibitem{Wu19}
Tingwei Wu, Dexin Qu, and Gengxin Zhang,
\newblock ``Research on {LoRa} adaptability in the {LEO} satellites internet of
  things,''
\newblock in {\em 2019 15th International Wireless Communications \& Mobile
  Computing Conference (IWCMC)}, 2019, pp. 131--135.

\end{thebibliography}

\end{document}